\documentclass[12pt]{article}
\usepackage{graphicx}
\usepackage{wrapft}
\usepackage{wrapfig}
\usepackage{cite}
\usepackage{amsmath}
\usepackage{amssymb}
\usepackage{amsfonts}
\usepackage{physics}
\usepackage{ulem}
\usepackage{color}
\usepackage{comment}
\numberwithin{equation}{section}
\def\be{\begin{eqnarray}}
\def\ee{\end{eqnarray}}
\def\eq{\label}
\def\nn{\nonumber  \\}
\def\abstract#1{\vskip 7mm 
\begin{center}{\large Abstract}\par \bigskip
\begin{minipage}[c]{12cm}
\small #1
\end{minipage}
\end{center}
}
\def\title#1{\begin{center}{\Large\bf #1}\end{center}}
\def\author#1{\vskip 5mm \begin{center}{#1}\end{center}}
\def\address#1{\begin{center}{\it #1}\end{center}}
\newcommand{\bfr}{\begin{flushright}}
\newcommand{\efr}{\end{flushright}}
\begin{document}
\title{Gauge Theoretical Method in Solving Zero-curvature Equations I \\[1ex]\large{---
Application to the Static Einstein-Maxwell Equations with Magnetic Charge---}}
\author{
T.~Azuma
\footnote{azuma@dokyo.ac.jp}
}
\address{Dokkyo University\\
1-1 Gakuencho, Soka, Saitama 340-0042, Japan 
}
\author{
T.~Koikawa
\footnote{koikawa@otsuma.ac.jp}
}
\address{Institute of Human Culture Studies, Otsuma Women's University\\
12 Sanban-cho, Chiyoda-ku, Tokyo 102-8357, Japan}
\hspace{0.54cm}
\vspace{4cm}
\abstract{
The inverse scattering problem is applied to 2-dimensional partial differential equations called soliton equations such as the KdV equation and so on. It is also used to integrate the Einstein equations with axial symmetry. These inverse scattering problems look different. We show that they can be understood in a unified way. As an application to the Einstein equation, we find  solutions of the Einstein-Maxwell equations with a magnetic charge.}
\thispagestyle{empty}
\newpage
\section{Introduction}
Soliton equations are a class of nonlinear partial differential (NLPD) equations which can be studied and solved by several methods. One of them is the method called the inverse scattering method (ISM), which reduces NLPD equations to simultaneous linear equations. The nonlinear Schr\"odinger (NLS) equation, the Korteweg-de Vries (KdV) equation and the Kadomtsev-Petviashvili equation are examples of soliton equations. Besides the soliton equations, the ISM is also applied to the integration of the Einstein equation. The ISM is used not only to find soliton solutions but also to find which NLPD equations are soliton equations. 

We recapitulate how the ISM is developed in the soliton equations. The Miura transformation\cite{miura} connects two nonlinear partial differential equations. Suppose that the variable $u$ satisfies the KdV equation. Define other variable $v$ by $u=v,_x+v^2$. Then we find that $v$ satisfies the Modified Korteweg-de Vries (MKdV) equation. We rewrite the Miura transformation by introducing $\phi$ by $v=\phi,_x/\phi$. Then we obtain $v,_x+v^2=\phi,_{xx}/\phi$. Making use of fact that the KdV equation is invariant under the Galilean transformation, we can change $u$ to $u+\lambda$ where $\lambda$ is a parameter. We thus obtain the Shr\"odinger equation which is well known in the quantum mechanics\cite{GGKM}:
\be
-\phi,_{xx}+u\phi=\lambda\phi.\label{SE}
\ee  
Here $\phi$ is the wave function and $u$ is the potential. When the potential is given we solve the Schr\"odinger equation and determine the eigenvalue $\lambda$ and the eigenfunction 
$\phi$. Inversely, when the eigenvalues and the information of the wave function are given, we can construct the potential. Note that the potential $u$ is the solution to the KdV equation and this means that constructing $u$ by means of ISM is nothing but solving the KdV equation. 

The ISM was first applied to the KdV equation, and it was found that the ISM is also applicable to the NLS equation\cite{ZakShab1,ZakShab2} and the MKdV equation\cite{Wadachi1,Wadachi2}. 
M.J Ablowitz, D.J. Kaup, A.C. Newell and H. Segur (AKNS) introduced $2\times 2$ matrices 
$\mathcal{P}$ and $\mathcal{Q}$, and showed that a lot of soliton equations are solved by the ISM\cite{akns}. Instead of the second order differentioal equation (\ref{SE}), the AKNS system is composed of two linear equations called a scattering problem and a time evolution equation:
\be
\phi,_x&=&\mathcal{P}\phi,\eq{scattering}\\ 
\phi,_t&=&\mathcal{Q}\phi.\eq{timeevolution}
\ee
The compatibility condition of Eqs.(\ref{scattering}) and (\ref{timeevolution}) is given by
\be
\mathcal{P},_t-\mathcal{Q},_x+[\mathcal{P}, \mathcal{Q}]=0.\eq{akns}
\ee
From this condition numerous soliton equations can be drived.
The $2\times 2$ matrix form potentials $\mathcal{P}$ and $\mathcal{Q}$ are generalized to $N\times N(N\ge 2)$ matrices to accommodate more soliton equations\cite{aks}  

The ISM is also applied to the Einstein equations. Belinski and Zakharov extended the ISM to solve the vacuum Einstein equation with the metric depending on two variables\cite{BZ1}
and obtained the Kerr-NUT solution in the stationary axisymmetric case\cite{BZ2}. This metric also includes various cases such as cosmological solutions, cylindrical solutions and solutions for colliding waves\cite{BV}. 

In this paper, we show that the ISM is formulated by a pair of linear equations for the various NLPD equations. The compatibility condition of the pair of linear equations gives rise to the soliton equations. The same idea also applies to the Einstein equation. We define the gauge transformation by using the pair of linear equations, which leave the soliton equation invariant under the gauge transformation. We give a unified viewpoint to the NLPD equations including the Einstein equation. 

In theoretical high energy physics, it is known that the theory governing four interactions; strong interaction, electromagnetic interaction, weak interaction and gravitational interaction is the gauge theory. The gauge theory is a field theory in which the Lagrangian is invariant under the local gauge transformation. The invariance of the Lagrangian is guaranteed by changing the derivatives to covariant derivatives. 
In order to explain the gauge theory we exemplify a $O(N)$ scalar field  model of which the Lagrangian density is given by
\be
\mathcal{L}=\frac{1}{2}(\partial^{\mu}\Psi)^T \partial_{\mu}\Psi,
\ee
where $\Psi$ is a $N$-component vector: $\Psi=(\Psi_1,\Psi_2,\cdots, \Psi_N)^T$.
Here $\Psi^T$ stands for the transposition of $\Psi$.
Suppose that $\Psi(x)$ undergoes the transformation
\be
\Psi\to \Psi'=M\Psi,
\ee
where $M$ is a $N\times N$ matrix. As for $M$, we discuss two cases.  The first case is that $M$ is a constant matrix. Second case is that $M$ is $x$-dependent: $M=M(x)$. If $M$ is a constant matrix, the Lagrangian density is invariant under the transformation as far as $M\in O(N)$. When $M$ is $x$-depnednt, the Lagrangian density is not invariant under the transformation because $\partial_\mu\Psi'=M(x)\partial_\mu\Psi+(\partial_\mu M(x))\Psi$. 
It is possible to restore the invariance of the Lagrangian density by replacing $\partial_{\mu}\Psi$ in the Lagrangian density by the covariant derivative $D_\mu\Psi=(\partial_{\mu}-A_\mu)\Psi$ and also imposing the gauge transformation on the gauge field $A_\mu(x)$ so that $D_\mu\Psi\to M(x)(D_\mu\Psi)$ under the gauge transformation. The Lagrangian deisity written in terms of the covariant derivatives becomes 
\be
\mathcal{L}=\frac{1}{2}(D^{\mu}\Psi)^T D_{\mu}\Psi.
\ee
The Euler-Lagrange equation derived from the Lagrangian density reads
\be
(D^\mu)^T D_\mu\Psi=0.\eq{gONeq}
\ee
This shows that the solution to the Euler-Lagrange equation is obtainable from the first order equation
\be
(\partial_\mu-A_\mu)\Psi=0.
\ee
Therefore we here consider the first order equation given by
\be
(\partial_i-\mathcal{A}_i)\Psi=0,~(i=1,2)
\ee
where $\mathcal{A}_i=\mathcal{A}_i(\lambda,x_1,x_2)$ is, for example, the Lie algebra valued function of $x_1$ and $x_2$. Then, anybody would find that the covariant derivatives have the same form as in Eqs.(\ref{scattering}) and (\ref{timeevolution}). Though the linear equations for the soliton case are almost the same as the covariant derivatives, the linear equations used in the Einstein equations look apparently different. In order to construct a theory which is applicable to the linear equations for the Einstein equation, we need to extend the ordinary covariant derivatives to more generalized covariant derivatives so that they are consistent with a coupled linear equation both for the soliton equations and the Einstein equation. Then we need to construct a unified formulation that is applicable both to the soliton equations and the Einstein equation. We call this theory the gauge theory from here on, for simplicity.

The ISM is to construct solutions or potentials from $\Psi$. In the gauge theory we transform $\Psi$ to $\Psi'$ by the gauge transformation $\Psi'=\mathcal{G}\Psi$, where $\mathcal{G}$ is a space dependent $N\times N$ matrix. Then a new solution obtained from $\Psi'$ by the ISM should also be a solution.
 However, the solution thus obtained is not always an interesting solution. The soliton solutions are characterized  by the soliton numbers. A simple gauge transformation does not yield a soliton solution with different soliton numbers. We show that the singular gauge transformation would give rise to the solutions with different soliton numbers about which we discuss for the Einstein equation.
 
 The terminology gauge transformation was used  by Wadati and Sogo\cite{ws} as the transformation that the AKNS system\cite{akns} and the WKI system\cite{WKI1,WKI2} are connected. The gauge transformation is regarded as an extension of the Miura transformation\cite{miura} connecting the KdV equation to the MKdV equation. This shows that their gauge transformation connects different soliton equations. 
On the other hand, the gauge transformation in the present paper connects a solution of a soliton equation with other solution of the same equation. Since the soliton solutions are characterized by the soliton numbers, we could mention that the gauge transformation connects a solution with a solution with different soliton number.

In the following section, we show the gauge theoretical formulation of solitons for both soliton equations and the Einstein equation. We also define the singular gauge transformation which gives rise to solutions with different soliton numbers.
In section 3, we show both differences and similarities between the inverse scattering method of soliton equations and the Einstein equation. Both fall into the general formulation given in section 2. 

There are many papers on the exact solutions of the Einstein-Maxwell equations. In these works, 
the applications of the ISM\cite{B,A} and of other solution-generating methods\cite{C,GE,NK} were presented. Using these methods, various soliton solutions have been studied\cite{DGN,GNG,AK1,RMM,AK2,BMS,MG,AB1,AK3,AK4,AB2,AB3}.
There are also works where  the solutions of the Einstein-Maxwell equations with a magnetic charge are constructed\cite{BO,BG,GG}. We are interested in finding that the solutions are like those of the Einstein-Maxwell equations with an electric charge. The Maxwell equation says that the total magnetic charge should be zero. This condition contradicts the existence of magnetic monopole.
This is resolved by a Dirac string which is compared to a thin solenoid in which magnetic flux flows exist. By introducing the Dirac string, the monopole has no contradiction with the Maxwell equation. It is not clear what it can be in the Einstein-Maxwell equations and what is its role in the gravitational field. In section 4, we apply the gauge theoretical method to solving the Einstein-Maxwell equations with a magnetic charge and discuss properties of the monopole black hole. In section 5, we give a brief discussion.  
\section{Formulation}
In this section, we will see that the soliton equations are expressed by the zero-curvature equation, and we discuss the gauge theory for soliton equations. 

Here a question arises. How can it be possible?  In the gauge theory we discuss the gauge invariance of Lagrangian while we do not use the Lagrangian for most of the soliton equations.

The zero-curvature equation arises as a compatibility condition of a pair of linear equations. Therefore, we will study the gauge transformation of the linear equations and the zero-curvature equation instead of considering the transformation of Lagrangian.

\subsection{Zero-curvature equation}
We discuss two linear equations given by
\be
\left(\hat D_1-\mathcal{A}_1(\lambda,x_1,x_2)\right)\Psi(\lambda,x_1,x_2)&=&0,\eq{linear1}\\
\left(\hat D_2-\mathcal{A}_2(\lambda,x_1,x_2)\right)\Psi(\lambda,x_1,x_2)&=&0.\eq{linear2}
\ee 
Here $\hat D_1$ and $\hat D_2$ are a pair of commuting differential operators, which are linear combinations of $\partial/\partial x_1$, $\partial/\partial x_2$ and possibly $\partial/\partial\lambda$ though $\lambda$ is set to be constant after the differentiation with respect to $\lambda$. $\mathcal{A}_i$ are $N\times N$ matrix-valued gauge potentials and $\hat D_i-\mathcal{A}_i,(i=1,2)$ are called covariant derivatives in the gauge theory. The variables $x_1$, $x_2$ could be time and one of two-space variables, or two-space variables, depending on the case we would apply to. $\Psi=\Psi(\lambda,x_1,x_2)$ is a complex $N\times N$ matrix function.

The compatibility condition of two equations (\ref{linear1}) and (\ref{linear2}) is given by
\be
\left[\hat D_1-\mathcal{A}_1,\hat D_2-\mathcal{A}_2\right]\Psi=0.
\ee
This leads to
\be
\left(\hat D_1\mathcal{A}_2-\hat D_2\mathcal{A}_1+[\mathcal{A}_2, \mathcal{A}_1]\right)\Psi=0.
\ee
If the nontrivial $\Psi$ should exist, we have
\be
\hat D_1\mathcal{A}_2-\hat D_2\mathcal{A}_1+[\mathcal{A}_2, \mathcal{A}_1]=0,\eq{zerocurvature}
\ee
which is called a zero-curvature equation. The zero-curvature equation leads to a soliton equation in the soliton case, and one of the equations for metric functions in the gravity case. 
\subsection{Gauge transformation for the zero-curvature equation}
We consider the local gauge transformation for two linear equations given in Eqs.(\ref{linear1}) and  (\ref{linear2}). 
Under the gauge transformation, we require that $\Psi$ and the gauge potentials $\mathcal{A}_i$ transform as
\be
\Psi &\to& \Psi'=\mathcal{G}\Psi,\cr
\mathcal{A}_i &\to& \mathcal{A}'_i=\mathcal{G} \mathcal{A}_i\mathcal{G}^{-1}-(\hat D_i\mathcal{G})\mathcal{G}^{-1},
\ee
where $\mathcal{G}=\mathcal{G}(\lambda,x_1,x_2)$ is a $N\times N$ matrix.
Then the gauge transformation for linear equations become 
\be
(\hat D_i-\mathcal{A}_i)\Psi&\to&(\hat D_i-\mathcal{A}'_i)\Psi'\cr
&=&(\hat D_i\mathcal{G})\Psi+\mathcal{G}(\hat D_i\Psi)-\left(\mathcal{G} \mathcal{A}_i\mathcal{G}^{-1}-(\hat D_i\mathcal{G})\mathcal{G}^{-1}\right)\mathcal{G}\Psi\cr
&=&\mathcal{G}(\hat D_i-\mathcal{A}_i)\Psi,\eq{gaugetrs}
\ee
which shows that the covariant derivative transforms in the same way as $\Psi$. Since
\be
(\hat D_i-\mathcal{A}'_i)\Psi'=\mathcal{G}(\hat D_i-\mathcal{A}_i)\Psi=\left(\mathcal{G}(\hat D_i-\mathcal{A}_i)\mathcal{G}^{-1}\right)\mathcal{G}\Psi,
\ee
we may write
\be
\hat D_i-\mathcal{A}'_i=\mathcal{G}(\hat D_i-\mathcal{A}_i)\mathcal{G}^{-1}.
\ee

Making use of the result, the zero-curvature equation transforms under the gauge transformation as
\be
&&[\hat D_1-\mathcal{A}_1, \hat D_2-\mathcal{A}_2]\Psi\cr
&\to&\left[\mathcal{G}(\hat D_1-\mathcal{A}_1)\mathcal{G}^{-1}, \mathcal{G}(\hat D_2-\mathcal{A}_2)\mathcal{G}^{-1}\right]\mathcal{G}\Psi\cr
&=&[\hat D_1-\mathcal{A}'_1, \hat D_2-\mathcal{A}'_2]\Psi'\cr
&=&\left(\hat D_1\mathcal{A}'_2-\hat D_2\mathcal{A}'_1+[\mathcal{A}'_2, \mathcal{A}'_1]\right)\Psi'.
\ee
Therefore when gauge potentials satisfy the zero-curvature equation, so do the  gauge potentials after the gauge transformation. 

As we mentioned before, the zero-curvature equation is the soliton equation, for example. This shows that starting with the trivial solution satisfied by potential $\mathcal{A}_i$, the gauge transformed potential $\mathcal{A}'_i$ are also the solution to the zero-curvature equation. Here arises a question. Are the new solutions interesting solutions? 

\subsection{Soliton solution by singular gauge transformation}

We should note that not all the solutions obtained by the gauge transformation are interesting solutions. It is well known that the soliton solutions are characterized by the soliton number. The new solution obtained by the regular gauge transformation does not change the soliton number. Starting with a trivial solution we obtain a non-trivial soliton solution by a singular gauge transformation. The singular gauge transformation means the transformation which is singular in $\lambda$ space.  The gauge transformation by $\mathcal{G}(\lambda,x_1,x_2)$ in Eq.(\ref{gaugetrs}) is singular in $\lambda$ space such that
\be
\mathcal{G}=1+\frac{R(x_1,x_2)}{\lambda-\mu(x_1,x_2)}.
\ee
The gauge transformation adds one soliton number. When we would like to add $N$ soliton number, this turns out to be
\be
\mathcal{G}=1+\sum_{i=1}^{N}\frac{R_i(x_1,x_2)}{\lambda-\mu_i(x_1,x_2)}\eq{Npgt}.
\ee

\section{Soliton equations and the Einstein equation}
\subsection{Soliton Equation}
We mentioned that soliton equations are written as the zero-curvature equations (\ref{zerocurvature}) of the formulation in the proceeding section. We show it explicitly in the following example.

We exemplify the NLS equation for the soliton equations, which is given by
\be
i\partial_t\phi=-\partial_x^2\phi-2|\phi|^2\phi,
\ee
where $\phi$ is a complex field. 

The variables for the soliton equations in the formulation are $x_1=x$ and $x_2=t$, and the differential operators in Eqs.(\ref{linear1}) and (\ref{linear2}) are given by
\be
\hat D_1&=&\partial_x,\\
\hat D_2&=&\partial_t.
\ee

The two linear equations are given by
\be
(\partial_x-\mathcal{A}_x)\Psi&=&0,\\
(\partial_t-\mathcal{A}_t)\Psi&=&0,
\ee
where the gauge potentials $\mathcal{A}_x$ and $\mathcal{A}_t$ are $su(2)$-valued functions given by
\be
\mathcal{A}_x&=&\mqty(-i\lambda & i\phi^* \\ i\phi & i\lambda)=i\{\phi_1\sigma_1+\phi_2\sigma_2-\lambda\sigma_3\},\\
\mathcal{A}_t&=&\mqty(2i\lambda^2-i|\phi|^2 & \phi^*,_x-2i\lambda\phi^* \\ -\phi,_x-2i\lambda\phi & -2i\lambda^2+i|\phi|^2)\cr
&=&i\left\{(-\phi_2,_x-2\lambda\phi_1)\sigma_1+(\phi_1,_x-2\lambda\phi_2)\sigma_2+(2\lambda^2-|\phi|^2)\sigma_3\right\}.
\ee
Here $\sigma_i$ are Pauli Matrices.

The compatibility condition of the above two equations read
\be
[\partial_x-\mathcal{A}_x,\partial_t-\mathcal{A}_t]\Psi=0,
\ee
which leads to the zero-curvature equation:
\be
\partial_x\mathcal{A}_t-\partial_t \mathcal{A}_x+[\mathcal{A}_t, \mathcal{A}_x]=0.
\ee
Substituting the gauge potentials here we obtain
\be
\mqty(0&-\partial_x^2\phi-i\partial_t\phi-2\phi|\phi|^2\\\partial_x^2\phi^*-i\partial_t\phi^*+2\phi^*|\phi|^2&0)=0,
\ee
which shows that the zero-curvature equation leads to the NLS equation.
\subsection{Vacuum Einstein equation with axial symmetry}
In the axially symmetric case, the metric is given by
\be
ds^2=F(\rho,z)(d\rho^2+dz^2)+g_{ab}(\rho,z)dx^adx^b,\quad (a,b=1,2)
\ee
where $x^1=t$ and $x^2=\phi$.
The vacuum Einstein equation with the coordinates condition $\det g=-\rho^2$ reads
\be
&&(\rho g,_\rho g^{-1}),_\rho+(\rho g,_zg^{-1}),_z=0,\label{eq1}\\
&&(\ln F),_\rho=-\frac{1}{\rho}+\frac{1}{4\rho}\Tr(U^2-V^2),\label{eq1Frho}\\
&&(\ln F),_z=\frac{1}{2\rho}\Tr(UV),\label{eq1Fz}
\ee
where $g$, $U$ and $V$ are the $2\times 2$ matrices defined by
\be
g=(g_{ab}),\quad U=\rho g,_\rho g^{-1},\quad V=\rho g,_zg^{-1}\label{UV}.
\ee
We find that the first equation (\ref{eq1}) is that for $g$ and the remaining equations (\ref{eq1Frho}) and (\ref{eq1Fz}) are those for $g$ and $F$. We need to solve the first equation and then solve the remaining equations by use of the solution for $g$. We focus on the first equation. We show how we get it as the zero-curvature equation.

In the present case, the two linear equations for the $2\times 2$ matrix function $\Psi(\lambda,\rho,z)$ are
\be
\hat D_1\Psi(\lambda,\rho,z)&=&\mathcal{A}_1\Psi(\lambda,\rho,z),\label{ldeq1}\\
\hat D_2\Psi(\lambda,\rho,z)&=&\mathcal{A}_2\Psi(\lambda,\rho,z),\label{ldeq2}
\ee
where
\be
\hat D_1&=&\partial_z-\frac{2\lambda^2}{\lambda^2+\rho^2}\partial_\lambda,\label{do1}\\
\hat D_2&=&\partial_\rho+\frac{2\lambda\rho}{\lambda^2+\rho^2}\partial_\lambda,\label{do2}
\ee
and
\be
\mathcal{A}_1&=&\frac{\rho V-\lambda U}{\lambda^2+\rho^2},\label{gp1}\\
\mathcal{A}_2&=&\frac{\rho U+\lambda V}{\lambda^2+\rho^2}.\label{gp2}
\ee
The compatibility condition
\be
[\hat D_1-\mathcal{A}_1, \hat D_2-\mathcal{A}_2]\Psi=0
\ee
leads to the zero-curvature equation (\ref{zerocurvature}) and it is explicitly expressed as
\be
\left(\rho^2\left\{\left(\frac{U}{\rho}\right),_z-\left(\frac{V}{\rho}\right),_\rho
+\left[\left(\frac{U}{\rho}\right), \left(\frac{V}{\rho}\right)\right]\right\}
+\lambda(U,_\rho+V,_z)\right)\Psi=0.
\ee
Requiring that this should hold regardless of the value of $\lambda$, we obtain
\be
\left(\frac{U}{\rho}\right),_z-\left(\frac{V}{\rho}\right),_\rho+\left[\left(\frac{U}{\rho}\right), \left(\frac{V}{\rho}\right)\right]&=&0,\eq{zceqinee}\\
U,_\rho+V,_z&=&0.\eq{UrhoVz}
\ee
Then we find that Eq.(\ref{zceqinee}) is the zero-curvature equation for 
$\mathcal{A}_1(\lambda=0)=g,_zg^{-1}$ and $\mathcal{A}_2(\lambda=0)=g,_\rho g^{-1}$, and the continuity-like equation (\ref{UrhoVz}) is reduced to Eq.(\ref{eq1}). We conclude that the zero-curvature equation is used both in deriving the first equation of the Einstein equation and in expressing the soliton equations.
\section{Application to the Einstein-Maxwell equations with magnetic charge}
In this section, we apply the gauge theoretical method to solving the static and axisymmetric Einstein-Maxwell equations with a magnetic charge. We are interested in the gravitational field created by the magnetic charge, especially in the field by two magnetic charges. It would be interesting where the Dirac strings are located and how the equilibrium state is realized.  

In the case of the vacuum Einstein equation mentioned in the previous section, starting with a trivial solution of Eq.(\ref{eq1}), e.g., the flat metric $g_0=\mathrm{diag}(-1,\rho^2)$, its corresponding potential $\mathcal{A}_i$ and wave function $\Psi$, we can obtain a new solution by the singular gauge transformation (\ref{Npgt}). It is known that in the case $N=2$ we have the $2$-soliton solution that corresponds to the Kerr-NUT solution\cite{BZ2}.

In order to obtain the solutions of the static and axisymmetric Einstein-Maxwell equation with a magnetic charge, let us consider the metric
\be
ds^2 =- fdt^2+f^{-1}[e^k(d\rho^2+dz^2) + \rho^2d\phi^2], \label{metric}
\ee
where $f$ and $k$ are functions of $\rho$ and $z$. The source-free Einstein-Maxwell equations are given by
\be
&&R_{\mu\nu}=2\left( F_\mu{}^\alpha F_{\nu\alpha}
-\frac{1}{4}g_{\mu\nu}F^{\alpha\beta}F_{\alpha\beta}\right),\label{EMeq}\\
&&F^{\mu\nu}{}_{;\nu}=0,\\
&&F_{\mu\nu;\lambda}+F_{\nu\lambda;\mu}+F_{\lambda\mu;\nu}=0,\\
&&F_{\mu\nu}=A_{\nu;\mu}-A_{\mu;\nu}.\label{Feq}
\ee
In the magnetostatic case, the equations are written as
\be
&&(\ln f),_{\rho\rho} + \rho^{-1}(\ln f),_{\rho} + (\ln f),_{zz}
 = 2 \rho^{-2}f(\psi,_\rho^2+\psi,_z^2),\label{eqlnf}\\
&&k,_\rho = {\rho \over 2}[(\ln f),_\rho^2-(\ln f),_z^2] 
 +2\rho^{-1}f(\psi,_\rho^2-\psi,_z^2), \label{eqkrho}\\
&&k,_z = \rho(\ln f),_\rho(\ln f),_z +4\rho^{-1}f\psi,_\rho\psi,_z,\label{eqkz} \\
&&\psi,_{\rho\rho} - \rho^{-1}\psi,_{\rho}+\psi,_{zz}
 =-[\psi,_{\rho}(\ln f),_{\rho} + \psi,_z(\ln f),_z],\label{eqpsi}\\
&&\psi,_{\rho z}=\psi,_{z\rho},\label{eqpsirhoz}
\ee
where we have set $A_t=A_\rho=A_z=0$ and written $A_\phi$ as  $A_\phi=\psi$. If we introduce a function $\tilde\chi(\rho,z)$ by
\be
&&\psi,_\rho=\rho f^{-1}\tilde\chi,_z,\\
&&\psi,_z=-\rho f^{-1}\tilde\chi,_\rho,
\ee
and assume the relation with a constant $c_m$:
\be
&&f=1+2c_m\tilde\chi+\tilde\chi^2,\label{ftildechi}
\ee
we then find that the Eqs.(\ref{eqlnf})-(\ref{eqpsirhoz}) are reduced to 
\be
&&\tilde\chi,_{\rho\rho}+\rho^{-1}\tilde\chi,_{\rho}+\tilde\chi,_{zz}
=2f^{-1}(\tilde\chi+c_m)(\tilde\chi,_\rho^2+\tilde\chi,_z^2),\label{eqtildechi}\\
&&k,_\rho=2\rho f^{-2}(c_m^2-1)(\tilde\chi,_\rho^2-\tilde\chi,_z^2),\label{eqkrhotilde}\\
&&k,_z=4\rho f^{-2}(c_m^2-1)\tilde\chi,_\rho\tilde\chi,_z.\label{eqkztilde}\\
&&\tilde\chi,_{\rho z}=\tilde\chi,_{z\rho}.
\ee

We here introduce a new function $\bar f(\rho,z)$ defined by
\be
\bar f=\frac{Q_m+(d+m)\tilde\chi}{Q_m-(d-m)\tilde\chi},
\ee
where $Q_m$, $m$ and $d$ are constants satisfying $m=c_mQ_m$ and $d^2=m^2-Q_m^2$.
We find that this relation reduces Eq.(\ref{eqtildechi}) to 
\be
(\ln\bar f),_{\rho\rho}+\rho^{-1}(\ln\bar f),_\rho+(\ln\bar f),_{zz}=0,\label{eqbarf}
\ee
and the $2\times 2$ matrix $\bar g(\rho,z)$
\be
\bar g=\mqty(-\bar f^{-1}&0\\0&\bar f\rho^2)
\ee
satisfies the equation
\be
(\rho\bar g,_\rho\bar g^{-1}),_\rho+(\rho\bar g,_z\bar g^{-1}),_z=0.
\ee
Therefore, if we define $\bar U$ and $\bar V$ using $\bar g$ instead of $g$ in Eq.(\ref{UV}),  and the corresponding potentials $\mathcal{\bar A}_i(i=1,2)$ using $\bar U$ and $\bar V$ instead of $U$ and $V$ in Eqs.(\ref{gp1}) and (\ref{gp2}), we have the zero-curvature equation
\be
\hat D_1\mathcal{\bar A}_2-\hat D_2\mathcal{\bar A}_1+[\mathcal{\bar A}_2, \mathcal{\bar A}_1]=0.\eq{zerocurvature1}
\ee
Then we can obtain the solution of $\bar g$ from a trivial solution, e.g., $\bar g=\mathrm{diag}(-1, \rho^2)$ by the singular gauge transformation with $x_1=\rho$ and $x_2=z$ in Eq.(\ref{Npgt}). 

As we restrict $\bar g$ to a diagonal matrix, we have a $N$-soliton solution for $\bar f$:
\be
\bar f=\prod_{k=1}^N\left(\frac{i\mu_k}{\rho}\right),\label{barfsol}
\ee
where $\mu_k\ (k=1,2,\cdots,N)$ are the functions of $\rho$ and $z$ satisfy the differential equations\cite{BZ2}
\be
\mu_k,_\rho-2\rho\mu_k(\mu_k^2+\rho^2)^{-1}=0,\quad\mu_k,_z+2\mu_k^2(\mu_k^2+\rho^2)^{-1}=0.\label{difeqmuk}
\ee
This soliton solution leads to the solution of $\tilde\chi$, and then we obtain the solutions for the metric functions $f$ and $e^k$:
\be
f&=&\frac{4d^2\prod\limits_k^N(i\mu_k)}{\rho^N
\left[(d+m)\rho^N+(d-m)\prod\limits_k^N(i\mu_k)\right]^2},\label{Nf}\\
e^k&=&\frac{\rho^{N^2/2}\prod\limits_{k>l}^N(\mu_k-\mu_l)^2}
{\prod\limits_k^N(\mu_k^2+\rho^2)\prod\limits_l^N\mu_l^{N-2}C^{(N)}}\label{Nek},\quad(N\ge 2)
\ee
and the solution for the magnetic potential $\psi$:
\be
\psi=\frac{Q_m}{2d}\left(c^{(N)}-\sum_k^N\mu_k+Nz\right)\label{Npsi}.
\ee
In these solutions $\mu_k$ are given  explicitly by the solutions of Eq.(\ref{difeqmuk}):
\be
\mu_k=w_k-z+(-1)^{k-1}\sqrt{(w_k-z)^2+\rho^2},\quad(k=1,2,\cdots,N)
\ee
where $w_k$ are constants. In Eq.(\ref{Nek}) $C^{(N)}$ is the constant given by
\be
C^{(N)}=\prod_{i>j}^{N/2}[4(w_{2i}-w_{2j})(w_{2i-1}-w_{2j-1})]^2,
\ee
and in Eq.(\ref{Npsi}) $c^{(N)}$ is an arbitrary constant.

We first investigate the $N=2$ case, i.e. the 2-soliton solution for $\bar f$:
\be
&&\bar f=-\frac{\mu_1\mu_2}{\rho^2},
\ee
with
\be
\left\{
\begin{array}{l}
\mu_1=-d-z+\sqrt{(d+z)^2+\rho^2}\\
\mu_2=d-z-\sqrt{(d-z)^2+\rho^2},
\end{array}
\right.
\ee
where we have set $w_1=-d$ and $w_2=d$.
The solutions for $f$, $k$, and $\psi$ are given by
\be
f&=&-\frac{4d^2\rho^2\mu_1\mu_2}{[(m+d)\rho^2+(m-d)\mu_1\mu_2]^2},\label{sol2f}\\
e^k&=&\frac{\rho^2(\mu_1-\mu_2)^2}{(\mu_1^2+\rho^2)(\mu_2^2+\rho^2)},\label{sol2k}\\
\psi&=&\frac{Q_m}{2d}(2d-\mu_1-\mu_2-2z).\label{sol2psi}
\ee
We here study the structure of the solutions given by Eqs.(\ref{sol2f})-(\ref{sol2psi}).
In the spatial infinity $\sqrt{\rho^2+z^2}\to\infty$, the solutions behave as
\be
f&\sim&1-\frac{2m}{\sqrt{\rho^2+z^2}},\\
e^k&\sim&1,\\
\psi&\sim&Q_m\left(1-\frac{z}{\sqrt{\rho^2+z^2}}\right),
\ee
and the magnetic components of field strength as
\be
B^\rho&=&F_{\phi z}/\sqrt{-g}\sim\frac{Q_m\rho}{(\rho^2+z^2)^{3/2}},\\
B^z&=&F_{\rho\phi}/\sqrt{-g}\sim\frac{Q_mz}{(\rho^2+z^2)^{3/2}}.
\ee
Dividing the symmetry axis defined by $\rho=0$ into three regions (Fig.1), we have the behaviors of the solutions on the axis:

\begin{wrapfigure}[]{r}{5cm}
\vspace{0.3cm}
\centering\includegraphics[height=5cm]{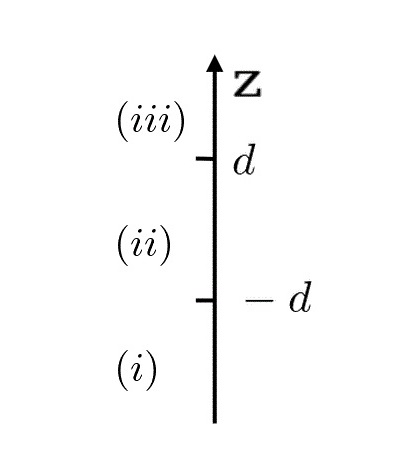}\\
Fig.1
\vspace{-3cm}
\end{wrapfigure}\noindent(i)$z<-d$ region
\be
f&=&\frac{z^2-d^2}{(z-m)^2},\\
e^k&=&1,\\
\psi&=&2Q_m,
\ee
\noindent(ii)$-d<z<d$ region
\be
f&=&0,\\
e^k&=&0,\\
\psi&=&Q_m\frac{d-z}{d},
\ee
\noindent(iii)$z>d$ region
\be
f&=&\frac{z^2-d^2}{(z+m)^2},\\
e^k&=&1,\\
\psi&=&0.
\ee
These behaviors show that there exists an event horizon in the region (ii) and the Dirac string in the region (i).

We shall express the two soliton solution in terms of the spherical coordinates.
In the spherical coordinates $(r,\theta)$ defined by
\be
&&\rho=\sqrt{(r-m)^2-d^2}\sin\theta,\\
&&z=(r-m)\cos\theta,
\ee
the solutions are expressed as
\be
f&=&1-\frac{2m}{r}+\frac{Q_m^2}{r^2},\\
\psi&=&Q_m(1-\cos\theta),
\ee
and the magnetic component of field strength as
\be
B^r&=&\frac{Q_m}{r^2}.
\ee
We find that the solution has the same form as the Reissner-Nordstr\"om solution, although it describes a monopole black hole with the Dirac string. The event horizons are located at $r=m\pm\sqrt{m^2-Q_m^2}$ and the Dirac string lies along the negative part of the $z$-axis.  Since we can choose the integration constant in the solution $\psi$ so that the string lies along  the positive part of the $z$-axis, we conclude that the Dirac string does not affect the spacetime structure. The quantity $\psi$ is not differentiable at the point $r=0$. Since one of the Maxwell equations showing that no monopole exists does not hold true at this point, the monopole become possible.

We calculate here the magnetic charge of monopole black hole and the masses of monopole black hole and Dirac string. The total magnetic charge and mass in the system described by the solutions are given by the integrals over the surface at spatial infinity $S_\infty$:
\be
Q_m^{\rm(total)}&=&\frac{1}{8\pi}\oint_{S_\infty}\epsilon^{lmn}F_{mn}d\sigma_l\nn
&=&\left[-\frac{1}{2}\psi(0,z)\right]^{z=+\infty}_{z=-\infty}\nn
&=&Q_m,\\
m^{\rm(total)}&=&\frac{1}{4\pi}\oint_{S_\infty}\xi^{t;l}\sqrt{-g}d\sigma_l\nn
&=&\left[-\frac{c}{2}\psi(0,z)\right]^{z=+\infty}_{z=-\infty}\nn
&=&m,\label{totalm}
\ee
where $\xi^t$ is the timelike Killing vector. The magnetic charge of monopole black hole can be defined by the integral over the suface at the horizon $S_H$:
\be
Q_m^{\rm(BH)}&=&\frac{1}{8\pi}\oint_{S_H}\epsilon^{lmn}F_{mn}d\sigma_l\nn
&=&\left[-\frac{1}{2}\psi(0,z)\right]^{z=+d}_{z=-d}\nn
&=&Q_m.
\ee
This shows that the magnetic charge is located inside the horizon. The mass of black hole can be also defined by the integral over $S_H$.
Taking account of the contribution of electromagnetic field to the integral\cite{BC}, we have
\be
m^{\rm(BH)}&=&\frac{1}{4\pi}\oint_{S_H}(\xi^{t;l}+\psi F^{\phi l})\sqrt{-g}d\sigma_l\nn
&=&\left[-\frac{1}{2}\psi(0,z)[c+\tilde\chi(0,z)]\right]^{z=+d}_{z=-d}\nn
&=&m-\frac{Q_m^2}{m+d}=m-\frac{Q_m^2}{m+\sqrt{m^2-Q_m^2}}.
\ee
The mass of Dirac string can be defined by the integral over the surface $S_{DS}$ around the Dirac string:
 \be
m^{\rm(DS)}&=&\frac{1}{4\pi}\oint_{S_{DS}}(\xi^{t;l}+\psi F^{\phi l})\sqrt{-g}d\sigma_l\nn
&=&\left[-\frac{1}{2}\psi(0,z)[c+\tilde\chi(0,z)]\right]^{z=-d}_{z=-\infty}\nn
&=&\frac{Q_m^2}{m+d}=\frac{Q_m^2}{m+\sqrt{m^2-Q_m^2}}.
\ee
We find that the total mass consists of the masses of black hole and Dirac string.
When $Q_m=0$ the Dirac string disappears and the solution becomes the Schwarzschild solution.

In the extremal case where $d\to 0$, the solutions in the $(r,\theta)$-coordinates become
\be
f&=&\left(1-\frac{|Q_m|}{r}\right)^2,\\
\psi&=&Q_m(1-\cos\theta).
\ee
The magnetic charge of black hole is given by  
\be
Q_m^{\rm(BH)}&=&Q_m,
\ee
however, the masses of black hole and Dirac string become
\be
m^{\rm(BH)}&=&0,\\
m^{\rm(DS)}&=&|Q_m|.
\ee
In this case, the Schwarzschild radius $r_g$ of black hole is given by the magnetic monopole charge:
\be
r_g=|Q_m|.
\ee

We next consider the $N=4$ case, i.e., the 4-soliton solution for $\bar f$. We study the behaviors of the solution and find where the Dirac string lies. We also study the individual magnetic charges and masses of two monopole black holes in the equilibrium case. The solution is given by
\be
\bar f&=&\frac{\mu_1\mu_2\mu_3\mu_4}{\rho^4}\label{4sbarf},
\ee
with
\be
\left\{
\begin{array}{l}
\mu_1=z_1-z-d_1+\sqrt{(z_1-z-d_1)^2+\rho^2},\\
\mu_2=z_1-z+d_1-\sqrt{(z_1-z+d_1)^2+\rho^2},\\
\mu_3=z_2-z-d_2+\sqrt{(z_2-z-d_2)^2+\rho^2},\\
\mu_4=z_2-z+d_2-\sqrt{(z_2-z+d_2)^2+\rho^2},
\end{array}
\right.
\ee
where $z_1$, $z_2$, $d_1$, $d_2$ are constants. From the solution (\ref{4sbarf}) we obtain the solutions for $f$, $Q$, and $\psi$:
\be
f&=&\frac{4d^2\rho^4\mu_1\mu_2\mu_3\mu_4}{[(m+d)\rho^4-(m-d)\mu_1\mu_2\mu_3\mu_4)]^2},
\label{4sf}\\
e^k&=&\frac{\rho^8[(\mu_1-\mu_2)(\mu_1-\mu_3)(\mu_1-\mu_4)(\mu_2-\mu_3)(\mu_2-\mu_4)(\mu_3-\mu_4)]^2}{(\mu_1^2+\rho^2)(\mu_2^2+\rho^2)(\mu_3^2+\rho^2)(\mu_4^2+\rho^2)(\mu_1\mu_2\mu_3\mu_4)^2C^{(4)}},
\label{4sk}\quad\quad \\
\psi&=&\frac{Q_m}{2d}[2(z_1+z_2+d_1-d_2)-\mu_1-\mu_2-\mu_3-\mu_4-4z],\label{4spsi}
\ee
where
\be
C^{(4)}=16[(z_1-z_2)^2-(d_1-d_2)^2]^2.\label{}
\ee
We first study the structure of the solutions given by Eqs.(\ref{4sf})-(\ref{4spsi}).
In the spatial infinity $\sqrt{\rho^2+z^2}\to\infty$, these solutions behave as
\be
f&\sim&1-\frac{2m(d_1+d_2)}{d\sqrt{\rho^2+z^2}},\label{4sfinf}\\
e^k&\sim&1,\label{4sQinf}\\
\psi&\sim&\frac{Q_m}{d}\left[d_1-d_2-\frac{(d_1+d_2)z}{\sqrt{\rho^2+z^2}}\right],\ \quad
\ee
and the magnetic components of field strength as
\be
B^\rho&\sim&\frac{Q_m(d_1+d_2)\rho}{d(\rho^2+z^2)^{3/2}},\\
B^z&\sim&\frac{Q_m(d_1+d_2)z}{d(\rho^2+z^2)^{3/2}}.
\ee
Hereafter we assume that $z_2+d_2>z_2-d_2>z_1+d_1>z_1-d_1$ without loss of generality (Fig.2). 
The behaviors of the solutions on the axis $\rho=0$ are as follows:

\noindent(i)$z<z_1-d_1$ region
\be
f&=&\frac{d^2(z-z_1-d_1)(z-z_1+d_1)(z-z_2+d_2)(z-z_2-d_2)}
{[d(z-z_1)(z-z_2)-md_2(z-z_1)-md_1(z-z_2)+dd_1d_2]^2},\quad\\
e^k&=&1,\\
\psi&=&\frac{2Q_md_1}{d},
\ee
\noindent(ii)$z_1-d_1<z<z_1+d_1$ region
\be
f&=&0,\\
e^k&=&0,\\
\psi&=&-\frac{Q_m(z-z_1-d_1)}{d},
\ee
\noindent(iii)$z_1+d_1<z<z_2-d_2$ region
\be
f&=&\frac{d^2(z-z_1-d_1)(z-z_1+d_1)(z-z_2+d_2)(z-z_2-d_2)}
{[d(z-z_1)(z-z_2)-md_2(z-z_1)+md_1(z-z_2)-dd_1d_2]^2},\quad\\
e^k&=&\left[\frac{(z_1-z_2+d_1+d_2)(z_1-z_2-d_1-d_2)}{(z_1-z_2+d_1-d_2)(z_1-z_2-d_1+d_2)}\right]^2,\label{regiiiek}\\
\psi&=&0,
\ee
\noindent(iv)$z_2-d_2<z<z_2+d_2$ region
\be
f&=&0,\\
e^k&=&0,\\
\psi&=&-\frac{Q_m(z-z_2+d_2)}{d},
\ee
\noindent(v)$z>z_2+d_2$ region
\be
f&=&\frac{d^2(z-z_1-d_1)(z-z_1+d_1)(z-z_2+d_2)(z-z_2-d_2)}
{[d(z-z_1)(z-z_2)+md_2(z-z_1)+md_1(z-z_2)+dd_1d_2]^2},\quad\\
e^k&=&1,\\
\psi&=&-\frac{2Q_md_2}{d}.
\ee
\begin{wrapfigure}[17]{r}{5cm}
\vspace{0.3cm}
\centering\includegraphics[height=7cm]{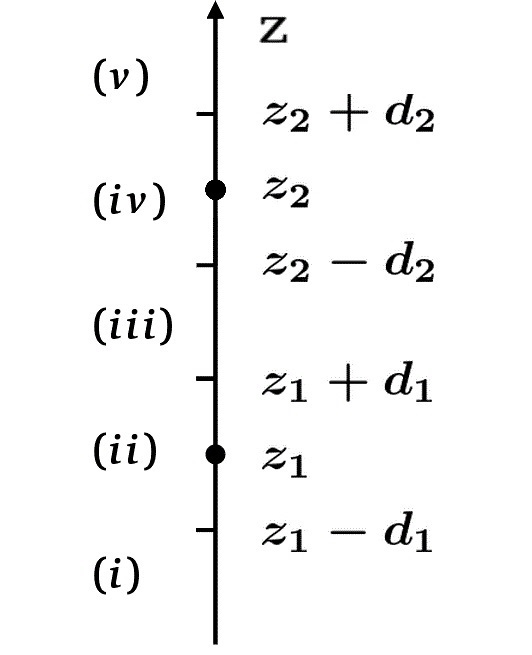}\\
Fig.2
\vspace{-4cm}
\end{wrapfigure}
From these behaviors of the solutions in the five regions, we find that the Dirac string lies in the regions (i)  and (v) along the $z$-axis. When we choose the integration constant in $\psi$ so that $\psi=0$ in the region (i), the Dirac string appears in the regions (iii) and (v). On the other hand, when we choose the constant so that $\psi=0$ in the region (v), the Dirac string appears in the regions (iii) and (v). The behavior of $e^k$ in the region (iii) shows that there is a strut in this region along the $z$-axis.

Last of all, we consider the equilibrium case in the solutions (\ref{4sf})-(\ref{4spsi}) where the strut in the region (iii) disappears. The equilibrium condition is given by
\be
\lim_{\rho\to 0}\left(\frac{g_{\phi\phi}}{\rho^2g_{\rho\rho}}\right)=\lim_{\rho\to 0}e^{-k}=1.
\ee
Imposing this condition on Eq.(\ref{regiiiek}), we have
\be
d_1d_2=0.\label{d1d20}
\ee 
In the case $d\ne 0$, the condition (\ref{d1d20}) leads to the one-monopole solution or the flat metric. In the case $d=0$, if we set $d_1=d_2=0$ keeping the ratios $d_1/d$ and $d_2/d$ finite, we have the equilibrium solution. By taking the limit $d_1, d_2, d\to 0$ in the solutions (\ref{4sf})-(\ref{4spsi}) and introducing the constants by
\be
Q_{m1}=\frac{Q_md_1}{d},\quad Q_{m2}=\frac{Q_md_2}{d},
\ee
we obtain
\be
f&=&\left(1+\frac{|Q_{m1}|}{r_1}+\frac{|Q_{m2}|}{r_2}\right)^{-2},\label{mpf}\\
e^k&=&1,\label{mpek}\\
\psi&=&\left(1-\frac{z-z_1}{r_1}\right)Q_{m1}-\left(1+\frac{z-z_2}{r_2}\right)Q_{m2},\label{mppsi}
\ee
where
\be
&&r_1=\sqrt{\rho^2+(z-z_1)^2},\qquad r_2=\sqrt{\rho^2+(z-z_2)^2}.
\ee
We find that the solutions in this case are of the Majumdar-Papapetrou type\cite{majpapa}.  They describe two monopole black holes with the Dirac strings stretching to $+\infty$ and $-\infty$ from each pole along the $z$-axis. As in the one-monopole blackhole case, the choice of the integration constant in $\psi$ does not affect the solutions Eqs.(\ref{mpf}) and (\ref{mpek}), 
the spacetime structure described by the solutions is the same as that investigated in detail in \cite{HH}.

Let us here study the individual charges and masses of the two monopole black holes described by the solutions (\ref{mpf})-(\ref{mppsi}). The total magnetic charge and mass are given by
\be
Q_m^{\rm(total)}&=&\frac{1}{8\pi}\oint_{S_\infty}\epsilon^{lmn}F_{mn}d\sigma_l\nn
&=&\left[-\frac{1}{2}\psi(0,z)\right]^{z=+\infty}_{z=-\infty}\nn
&=&Q_{m1}+Q_{m2},\\
m^{\rm(total)}&=&\frac{1}{4\pi}\oint_{S_\infty}\xi^{t;l}\sqrt{-g}d\sigma_l\nn
&=&\left[-\frac{c}{2}\psi(0,z)\right]^{z=+\infty}_{z=-\infty}\nn
&=&|Q_{m1}|+|Q_{m2}|.\label{}
\ee
The magnetic charges of the lower and upper black holes are defined by
\be
Q_m^{\rm(BH1)}&=&\frac{1}{8\pi}\oint_{S_{H1}}\epsilon^{lmn}F_{mn}d\sigma_l\nn
&=&\left[-\frac{1}{2}\psi(0,z)\right]^{z=z_1+0}_{z=z_1-0}=Q_{m1},\\
Q_m^{\rm(BH2)}&=&\frac{1}{8\pi}\oint_{S_{H2}}\epsilon^{lmn}F_{mn}d\sigma_l\nn
&=&\left[-\frac{1}{2}\psi(0,z)\right]^{z=z_2+0}_{z=z_2-0}=Q_{m2},
\ee
respectively. Here $S_{H1}$ and $S_{H2}$ are the event horizons of the lower and upper black holes, respectively. The masses of the lower and upper black holes are defined by
\be
m^{\rm(BH1)}&=&\frac{1}{4\pi}\oint_{S_{H1}}(\xi^{t;l}+\psi F^{\phi l})\sqrt{-g}d\sigma_l\nn
&=&\left[-\frac{1}{2}\psi(0,z)[c+\tilde\chi(0,z)]\right]^{z=z_1+0}_{z=z_1-0}\nn
&=&0,\\
m^{\rm(BH2)}&=&\frac{1}{4\pi}\oint_{S_{H2}}(\xi^{t;l}+\psi F^{\phi l})\sqrt{-g}d\sigma_l\nn
&=&\left[-\frac{1}{2}\psi(0,z)[c+\tilde\chi(0,z)]\right]^{z=z_2+0}_{z=z_2-0}\nn
&=&0,
\ee
respectively. The masses of the lower and upper Dirac strings are defined by
\be
m^{\rm(DS1)}&=&\frac{1}{4\pi}\oint_{S_{H1}}(\xi^{t;l}+\psi F^{\phi l})\sqrt{-g}d\sigma_l\nn
&=&\left[-\frac{1}{2}\psi(0,z)[c+\tilde\chi(0,z)]\right]^{z=z_1-0}_{z=-\infty}\nn
&=&|Q_{m1}|,
\ee
\be
m^{\rm(DS2)}&=&\frac{1}{4\pi}\oint_{S_{H2}}(\xi^{t;l}+\psi F^{\phi l})\sqrt{-g}d\sigma_l\nn
&=&\left[-\frac{1}{2}\psi(0,z)[c+\tilde\chi(0,z)]\right]^{z=+\infty}_{z=z_2+0}\nn
&=&|Q_{m2}|,
\ee
respectively.
From these calculations, we find that the total mass of this system is created by the magnetic charges and the magnetic charges give rise to the event horizons of monopole black holes. 

\section{Discussion}

In this section, we summarize what we did in this paper, and discuss what we are going to do following this paper.

One of the methods of obtaining exact solutions of the nonlinear partial differential equations called soliton is the inverse scattering method (ISM), which is also known to show what the soliton equations are. The inverse scattering method is also applied to the Einstein equation, though the ISM in the soliton equations and the Einstein equation look a little different. In high energy physics, the gauge theory is a central theory that is used in explaining the standard theory. Though the gauge theory focuses on the invariance of the Lagrangian, it is not used in the soliton equations in most of the cases. To apply the gauge theory to the soliton equations and the Einstein equation, we consider the invariance of a couple of linear equations and their compatibility condition, or zero-curvature equation. The gauge potentials in the zero-curvature equation satisfy the soliton equations. The invariance of the zero-curvature equation suggests that gauge transformation gives rise to new solutions by the gauge transformation. We show the formulation of the soliton equations and the Einstein equation and the method of obtaining soliton solutions by the gauge theory.

The concrete applications of the formulation to the soliton equations and the Einstein equation are shown. The Einstein-Maxwell equations with a magnetic charge with either the south pole or north pole are discussed. As for the Einstein-Maxwell equations with an electric charge, we have discussed it in Ref.\cite{AK1}. Here we discuss a black hole with a monopole with a magnetic charge with either north pole or south pole. A difference between the black hole with an electric charge and a magnetic charge is that a Dirac string should be inevitable in the discussion of Einstein-Maxwell equations with a magnetic charge.

As in the electric black hole, we construct soliton solutions with soliton numbers 2 and 4, corresponding to one monopole black hole and two monopole black holes respectively. The asymptotic behavior of the metric and the magnetic components in the 2-soliton solution is studied. Also, the behavior of the metric components on the symmetry axis defined by $\rho=0$ are studied, and the position of the event horizon and that of the Dirac string are shown there. The difference from the electric black hole is mentioned. The magnetic charge of the magnetic monopole and the masses of the magnetic monopole and the Dirac string are computed by the surface integral at spatial infinity and the event horizon.
We find that the mass of the black hole is equal to the total mass subtracted by that of the Dirac string. The asymptotic behavior of the metric and the magnetic components in the 4-soliton solution is studied. Also, the behaviors of the metric components in five regions defined on the symmetry axis are studied, and the positions of the Dirac string is identified. 

Since the 4-soliton solution represents the two-body system of magnetically charged black holes, we are interested in the equilibrium case where the strut disappears. This case corresponds to the Majumdar-Papapetrou type. In this case, one Dirac string stretches from $-\infty$ to the lower monopole black hole and another from the upper monopole black hole to $\infty$. The total charge and mass are computed. We find that the total mass of this system is created by the magnetic charges and the magnetic charges give rise to the event horizons of monopole black holes.

In a subsequent paper we are going to present the non-Weyl class solution of the Einstein-Maxwell equations with a magnetic charge. We apply the gauge theoretical method to solving the equations and obtain the solution where the gravitational and magnetic potentials are not functionally related. We find that the solution describes a magnetic dipole rather than a monopole in the gravitational field. 

\end{document}